\documentstyle[12pt]{article}
\newcommand{\D}{\displaystyle}

\newcommand{\h}{\hbar}
\newcommand{\pa}{\partial}
\newcommand{\var}{\varepsilon}

\begin{document}

\title{The Fundamental Commutator For Massless Particles}
%\begin{center}
%{\bf The Fundamental Comutator For Massless Particles}\\
%\vspace{0.5cm}
\author{$H. Razmi ^1\;\;\; \& \;\;\; Amir H. Abbassi ^2$}
%{\bf $H. Razmi^1$ and $A.H. Abbassi^2$ }
\maketitle
%\vspace{0.5cm}
\begin{center}
{\it \small Department of Physics, School of sciences \\
Tarbiat Modarres University, P.O.Box 14155-4838\\
Tehran, I.R. Iran }\\
$^1$\footnotesize  Email: razmi\_ha@net1cs.modares.ac.ir\\
$^2$ \footnotesize Email: ahabbasi@net1cs.modares.ac.ir\\
\end{center}  
\begin{abstract}
 It is discussed that the usual Heisenberg
commutation relation (CR) is not a proper relation for massless particles 
and then an alternative is obtained. The canonical quantization
of the free electromagnetic (EM)
fields based on the field theoretical generalization of this alternative
is carried out. Without imposing the normal ordering condition,
the vacuum energy is automatically zero. This can be considered as a solution
to the EM fields vacuum catastrophe and a step toward managing the cosmological constant problem at least for the EM fields contribution to
the state of vacuum.\\

\end{abstract}

\newpage
\subsection*{1. Introduction}
Quantum mechanically the relation $[\hat{Q}_i,\hat{P}_j]=i{\hbar} \delta_{ij}\hat{1}$ is generally 
accepted as the canonical CR between position and momentum 
operators of any particle regardless of its individual properties. For massive particles, this can be derived by means 
of symmetry group transformations and the related Lie algebra commutators [1-3]. We assume the position operator for the particle to be $\hat{Q}_i$ where 
by definition 
\begin{equation}
\hat{Q}_i \mid \vec q>=q_i\mid \vec q> \;\;\;\;\;\; i=1,2,3
\end{equation}\label{1}
\noindent has an unbounded continuous spectrum. Two assumptions are involved here, the 
space is a continuum and all three components of the position operator are
mutually commutative and so possess a common set of eigenvectors. The space displacement of localized position eigenvectors by applying the operator ,  $(e^{-i\frac {{\epsilon}_j}{\hbar}\hat{P_j}})$,
requires the displaced observables bear the same relationship to the displaced vectors as the original observables do to the original vectors, in particular \\
\begin{equation}
\hat Q_i'=e^{i\frac {{\epsilon}_j}{\hbar}\hat{P_j}}\hat Q_ie^{-i\frac {{\epsilon}_j}{\hbar}\hat {P_j}}
\end{equation} \label{2}
\noindent Up to first order approximation  in $\epsilon$, we may write
\begin{equation}
\hat Q_i'=(\hat 1+i\frac {{\epsilon}_j}{\hbar}\hat P_j)\hat Q_i(\hat 1-i\frac{{\epsilon}_j}{\hbar}\hat P_j)
\end{equation}   \label{3}
or
\begin{equation}
\hat Q_i'=\hat Q_i-i\frac{{\epsilon}_j}{\hbar}[\hat Q_i,\hat P_j]
\end{equation}  \label{4}   
\noindent By the substitution of  $[\hat{Q}_i,\hat{P}_j]=i\hbar {\delta}_{ij}$, in (4) one gets \\
\begin{equation}
\hat{Q}_i'=\hat Q_i+{\epsilon}_i\hat 1
\end{equation}  \label{5}
\noindent which means translation of reference frame is equivalent to translation of the particle.  This equivalence has route in this fact  that for a massive particle one can always transform to its rest frame.
However for massless particles e.g. photons, this transformation is impossible.
It is a property of Poincar\'e 
transformations that there is no rest frame for massless particles,they move with constant velocity in any reference frame.
Consequently, the canonical CR between coordinates and momenta is not
the same as massive for massless particles and we should look for the proper
relation.
\subsection*{2. Alternative Canonical CR For Massless Particles} 
 According to the Hamilton's equation of motion we can write
\begin{equation}
\frac {dQ_i}{dt}=\frac{\partial H}{\partial P_i}=\frac {\partial \sqrt {{P}^2{C}^2+m^2}}{\partial P_i}={H}^{-1}P_i
\end{equation}\label{6}
\noindent On the other hand by the Poisson brackets formalism, we have
\begin{equation}
\frac {dQ_i}{dt}= [Q_i , H]_p 
\end{equation} \label{7}
\noindent Therefore, classically (6) and (7) give
\begin{equation}
H^{-1}P_i = [Q_i, H]_p
\end{equation}  \label{8}
\noindent where $[ , ]_p$ represents the classical Poisson bracket.\\
Now, by means of Dirac's canonical quantization rule $[ , ] \rightarrow \frac {[ , ]}{i\hbar}$
we have the following operatorial relation
\begin{equation}
\hat {H^{-1}}\hat {P_i} = \frac {[\hat {Q_i} , \hat {H}]}{i\hbar}
\end{equation}  \label{9}
\noindent where $\hat{H}$ and $\hat{P}$ commute with each other.
By means of the Heisenberg equation of 
motion for position operator based on the fact that the Hamiltonian operator 
is the generator of time translation, we have 
\begin{equation}
[\hat{Q_i}, \hat{H}]=i{\hbar} \frac{d\hat{Q}_i}{dt}
\end{equation} \label{10}
\noindent Comparision of (9) and (10) gives 
\begin{equation}
\hat{P}_i= \left(\frac{\hat{H}}{c^2} \right )
\left(\frac{d\hat{Q}_i}{dt} \right )
\end{equation} \label{11}
\noindent Summing over square form of
(11) leads to $\hat{H}^2=\hat{P}^2$ which is an accepted relation for massless
particles.
 Using (11), we can write the commutator of 
$\hat{Q}_i$ and $\hat{P}_j$ as
\begin{equation}
[\hat{Q}_i, \hat{P}_j] =\\
\left[ \hat{Q}_i, \frac{\hat{H}}{c^2} \frac{d\hat{Q}_i}{dt} \right]=\\
\frac{1}{c^2} [\hat{Q}_i, \hat{H}] \frac{d\hat{Q}_j}{dt}+\frac{\hat{H}}{c^2} 
\left[ \hat{Q}_i, \frac{d\hat{Q}_j}{dt} \right] \\
\end{equation} \label{12}
 \noindent Insertion of (10) in (12) gives,
\begin{equation}
[\hat{Q}_i, \hat{P}_j] =\\
\frac{i{\hbar}}{c^2} \left( \frac{d\hat{Q}_i}{dt}\right)
\left( \frac{d\hat{Q}_j}{dt}\right) + \frac{\hat{H}}{c^2} 
\left[ \hat{Q}_i, \frac{d\hat{Q}_j}{dt} \right] 
\end{equation} \label{13}
For massless particles, we have
\begin{equation}
\begin{array}{lll}
\D{\sum_{j=1}^3} \left(\frac{d\hat {Q_j}}{dt} \right)^2 =c^2 \hat{1}   
\hspace*{2cm}\\
\D{\frac{d^2 \hat {Q_j}}{dt^2}=0}  
\end{array}
\end{equation} \label{14}
\noindent Then from (14) we may infer that
\begin{equation} 
\sum_{j=1}^3 \left [ \hat{Q}_i, 
\left ( \frac{d\hat{Q}_j}{dt} \right)^2 \right]=0 
\Rightarrow \sum_{j=1}^3 (\left [ \hat{Q}_i, 
 \frac{d\hat{Q}_j}{dt} \right]\frac{d\hat{Q}_j}{dt}+\frac{d\hat{Q}_j}{dt} 
\left[ \hat{Q}_i, \frac{d\hat{Q}_j}{dt} \right])=0 
\end{equation}  \label{15}
\noindent But we have
\begin{equation}
\frac{d}{dt} \left [ \hat{Q}_i, \frac{d\hat{Q}_j}{dt} \right] =
\left [ \frac{d\hat{Q}_i}{dt}, \frac{d\hat{Q}_j}{dt} \right]+ 
 \left [ \hat{Q}_i, \frac{d^2\hat{Q}_j}{dt^2} \right] = [\hat{H}^{-1} 
\hat{P}_i, \hat{H}^{-1} \hat{P}_j ]=0
\end{equation}  \label{16}
Therefore $\left[ \hat{Q}_i, \frac{d\hat{Q}_j}{dt} \right]$ is a constant 
operator or at most a function of the operators $\hat{H}$ and $\hat{P}$.
  Since $\frac{d\hat{Q}_j}{dt}=\hat{H}^{-1} \hat{P}_j$, we will have 
\begin{equation}
\left[ [\hat{Q}_i, \frac{d\hat{Q}_j}{dt} ] , \frac{d\hat{Q}_j}{dt} \right] =0
\end{equation} \label{17} 
\noindent Using (17) and (15)  will give
\begin{equation}
\D{\sum_{j=1}^3} \left ( \left [ \hat{Q}_i, \frac{d\hat{Q}_j}{dt} \right] 
\frac{d\hat{Q}_j}{dt} \right) =0
\end{equation} \label{18}
(18) may be written in the form
\begin{equation}
\hat{Q}_{i} \sum_{j} \left ( \frac{d\hat{Q}_j}{dt} \right )^2 -
\sum_{j} \frac{d\hat{Q}_j}{dt} \hat{Q}_i \frac{d\hat{Q}_j}{dt} =0
\end{equation}
By using (14), (19) gives
\begin{equation}
\hat{Q}_i = \frac 1{c^2} \sum_{j} \frac {d\hat{Q}_j}{dt} \hat{Q}_i
\frac {d\hat{Q}_j}{dt}
\end{equation}
$\{ \mid \vec k >\}$ is the complete set of eigenvectors of $\frac{d\hat{Q}_j}
{dt}$ ,
 \begin{equation}
\frac{d\hat{Q}_j}{dt}|\vec k>=c\frac{k_j}{|\vec k|}|\vec k>
\end{equation} \label{19}
The matrix element of (20) between $<\vec {k^\prime} \mid$ and $\mid
\vec k >$ gives 
\begin{equation}
<\vec {k^\prime}\mid \hat{Q}_i \mid\vec k >=
<\vec {k^\prime}\mid\hat{Q}_i \mid \vec k >\sum_{j} \frac {k_j 
{k^\prime}_j}{{\mid \vec k\mid}^2}
\end{equation}
(22) holds if and only if $<\vec {k^\prime}\mid \hat{Q}_i \mid\vec k > =0$
for $\vec {k^\prime} \not = \vec k$.\\
Thus there is only diagonal elements which means the complete set $\{\mid\vec k >\}$ is simultaneous
eigenvectors of $\hat{Q}_i$. Since $\hat{Q}_i$ and $\frac{d\hat{Q}_j}{dt}$
are self-adjoint and have a common complete set of eigenvectors we have 
\begin{equation}
\left[ \hat{Q}_i, \frac{d\hat{Q}_j}{dt} \right]=0
\end{equation}  \label{23}
Finally the desired CR for massless 
particles may be obtained by inserting (23) in (13). 
\begin{equation}
[\hat{Q}_i,\hat{P}_j]= \frac{i{\hbar}}{c^2} 
\left( \frac{d\hat{Q}_i}{dt}\right) 
\left( \frac{d\hat{Q}_j}{dt}\right) 
\end{equation}  \label{24}
It should be  noticed that if there were a rest frame for massless
particles, then $\frac{dQ_i}{dt}$, $\frac{dQ_j}{dt}$, and $c^2$ could
be set to go to zero and the right-hand side of (24) became $i\hbar {\delta}_{ij}$ (The limit of $\frac{a_ia_j}{a^2}$ as $a_i, a_j, a^2 \rightarrow 0$
is the Kronecker delta).\\
By means of (11), (24) can be also written in the following form
\begin{equation}
[\hat{Q}_i,\hat{P}_j]=i{\hbar} c^2 H^{-2} \hat{P}_i \hat{P}_j
\end{equation} \label{25}
Another possible form for this CR is as follows. Let's consider the eigenvalue
equation for momentum operator $\hat {\vec P}$
\begin{equation}
\hat {\vec P}\mid \vec k> = \hbar \vec k\mid \vec k>
\end{equation} \label{26}
\noindent Since $\hat H$ commutes with $\hat {\vec P}$, for free
particles, the eigenvectors $\mid \vec k>$'s are simultaneous eigenvectors of
$\hat H$ with the eigenvalue equation
\begin{equation}
\hat H\mid \vec k> = \hbar \omega \mid \vec k>
\end{equation} \label{27}
\noindent where, $\omega = c|\vec k|$.\\
Now, by means of the identity operator $\D{\sum_{\vec k}}\mid \vec k><\vec k\mid = \hat 1$,
we can write
\begin{equation}
[\hat Q_i,\hat P_j]=i\hbar c^2\D{\sum_{\vec k}}\D{\sum_{\vec k'}}\mid \vec k><\vec k\mid \hat {H^{-2}}\hat P_i\hat P_j\mid \vec k'><\vec k'\mid
\end{equation}  \label{28}
Using (26),(27), and the orthonormality of eigenvectors $\mid \vec k>$, (28) becomes 
\begin{equation}
[\hat Q_i,\hat P_j]=i\hbar\D{\sum_{\vec k}}\frac {k_ik_j}{{|\vec k|}^2}\mid \vec k><\vec k\mid
\end{equation} \label{29}
If one were able to find a frame in which $\vec k = 0$ (the rest frame of the particle),
then the right-hand side of (29) would become $i\hbar{\delta}_{ij}\hat 1$
which is the common result. But since there is no such rest frame for massless
particles we should work with the new proposed CR or its equivalence when
dealing with such particles.\\
It is evident  from (26) and (27) that
for any physical state whose energy-eigenvalue is zero, automatically its
momentum-eigenvalue is zero too; this prevents the singularity in
(25) due to the appearence of the operator ${\hat H}^{-2}$.\\
\subsection*{3. The Canonical Quantization of the Free EM Fields} 
For free EM fields we have;
\begin{equation}
L=-\frac{1}{4} F_{\mu \nu} F^{\mu\nu}
\end{equation} \label{30}
\begin{equation}
F^{\mu \nu}=\partial^{\nu}A^{\mu}-\partial^{\mu}A^{\nu}
\end{equation}  \label{31}
\begin{equation}
\Box A^{\mu}-\partial^{\mu} (\partial_{\nu} A^{\nu})=0
\end{equation} \label{32}
In Lorentz gauge, Fermi Lagrangian density is 
\begin{equation}
L_f=-\frac{1}{2} (\pa_{\nu}A_{\mu})(\pa^{\nu} A^{\mu})\hspace*{2cm}
\end{equation} \label{33}
\noindent then (32) gives
\begin{equation}
\Box A^{\mu}=0
\end{equation} \label{34}
\noindent and 
\begin{equation}
\pi_{\mu}=\frac{\pa L}{\pa \dot{A}^{\mu}}=-\frac{1}{c^2} \dot{A}_{\mu}
\end{equation} \label{35}
\noindent where ${\pi}_{\mu}$ is the canonical momentum operator conjugated to the field operator
four-potential $A_{\mu}$.
The solutions of the equation $\Box A^{\mu}=0$ can be expanded in terms of a complete 
set of solutions of the wave equation. Fourier expansion by imposing the 
periodic boundary condition gives 
\begin{equation}
A^{\mu} (x)=A^{\mu^+}(x) +  A^{\mu^-} (x)
\end{equation} \label{36}
where
\begin{equation}
A^{\mu^+}(x)=\D{\sum_r}\D{\sum_{\vec k}} 
\left( \frac{\h c^2}{2V \omega} \right)^{\frac{1}{2}} \var^{\mu}_r (\vec k) 
\hat{a}_r (\vec k) e^{-ikx}
\end{equation} \label{37}
\begin{equation}
A^{\mu^-}(x)=\D{\sum_r}\D{\sum_{\vec k}} 
\left( \frac{\h c^2}{2V \omega} \right)^{\frac{1}{2}} \var^{\mu}_r (\vec k) 
\hat{a}_r^+ (\vec k) e^{+ikx}
\end{equation} \label{38}
,$\omega=ck^0 =c|\vec k|$ and $A^{\mu}$ is appropriately normalized.\\
The summation over $r$, from $r=0$ to $r=3$, corresponds to the fact that for 
the field $A^{\mu}(x)$ there exists four linearly independent 
polarization states for each $\vec k$. These are described by the polarization 
vectors $\var_r^{\mu} (\vec k)$ which we choose to be real and satisfy the 
orthonormality and completeness relations
\begin{equation}
\var_r(\vec k)\var_s(\vec k)=\var_{r \mu} (\vec k) \var_s^{\mu} (\vec k)=-\xi_r\delta_{rs}
\end{equation} \label{39}
\begin{equation}
\D{\sum_r} \xi_r \var_r^{\mu} (\vec k) \var_r^{\nu} (\vec k) =-g^{\mu \nu}
\end{equation} \label{40}
\begin{equation}
\left \{ \begin{array}{l}
\xi_0=-1,\xi_1=\xi_2=\xi_3=1\\
g_{00}=-g_{11}=-g_{22}=-g_{33}=1
\end{array} \right .
\end{equation}  \label{41}
Construction of the canonical quantum field theories are based on the
equal-time commutation relations of the field operators and their
canonically conjugated momenta. For the free EM fields, the following equal-time commutation relation
has been generally accepted and used
\begin{equation}
[A^{\mu}(\vec x, t), {\pi}^{\nu}(\vec {x'}, t)]=-i\hbar c^2g^{\mu \nu}\delta(\vec x-\vec {x'})
\end{equation}  \label{42}
Indeed, this relation is the field theoretical, continuous, generalization of
the usual canonical commutation relation $[\hat {Q}_i,\hat {P}_j]=i\hbar \delta_{ij}\hat 1$
between the conjugate coordinates and momenta of the discrete
lattice approximation. But since we have already obtained an alternative CR for massless particles,
we should modify the relation (42) and find the appropriate corresponding
relation.
Following the same procedure generally used to pass from quantum mechanics
of discrete systems to canonical quantum field theories and by means of (29),
we may write
\begin{equation}
\left[ \hat{A}^{\mu} (\vec x,t),\hat{\pi}^{\nu}(\vec x',t)\right]=i\hbar (\D{\sum_{\vec K}}\frac {K^{\mu}K^{\nu}}{{K_0}^2}\mid \vec K><\vec K\mid) \delta(\vec x-\vec x')
\end{equation} \label{43}
\noindent in which $K_0=|\vec K|$.
If there were a rest frame for the system, the expression $\frac {K^{\mu}K^{\nu}}{{K_0}^2}$
could be set equal to $-g^{\mu \nu}$ and the relation (42) could be derived.
Of course, there is not such a rest frame.
After finding the proper CRs between $\hat{a}_r$ and 
$\hat{a}_r^+$'s, one can show that the relation (43) preserves the local 
property of the theory and the microcausality relation can be verified (see
Appendix -A). \\
Let us verify what the relation (43) gives for the CR between
$\hat{a}_r$ and \\
$\hat{a}_r^+$. Using (35), (43) becomes 
\begin{equation}
-\frac{1}{c^2} \left [\hat{A}^{\mu} (\vec x,t),\dot{\hat{A}}^{\nu}(\vec x',t) \right]=i\hbar (\D{\sum_{\vec K}}\frac {K^{\mu}K^{\nu}}{{K_0}^2}\mid \vec K><\vec K\mid) \delta(\vec x-\vec x')
\end{equation} \label{44}
or 
\begin{equation}
-\frac{1}{c^2} \D{\sum_r} \D{\sum_{r'}} \D{\sum_{\vec k}} \D{\sum_{\vec k'}} 
\left[ \frac{\h c^2}{2V \omega} \right]^{\frac{1}{2}}
\left[ \frac{\h c^2}{2V \omega'} \right]^{\frac{1}{2}}
\var_r^{\mu} (\vec k)\var_{r'}^{\nu} (\vec k') [\hat{a}_r(\vec k) e^{-ikx} +\hat{a}_r^+
(\vec k) e^{ikx},
\end{equation} \label{45}
\begin{equation}
(-ik{'}^{0} 
 c)\hat{a}_{r'} (\vec k') e^{-ik' x'}+(ik{'}^{0} c)\hat{a}_{r'}^+ (\vec k') 
e^{ik'x'}]=i\hbar (\D{\sum_{\vec K}}\frac {K^{\mu}K^{\nu}}{{K_0}^2}\mid \vec K><\vec K\mid) \delta(\vec x-\vec x')
\end{equation} \label{46}
It should be noticed that $t=t'$ and $k^0 c=\omega$. \\
The application of $\D{\sum_{\vec k}} \frac{1}{\sqrt{V}} e^{i\vec k\cdot 
 \vec x} \frac{1}{\sqrt{V}} e^{-i\vec k \cdot 
\vec x'}=\delta (\vec x-\vec x')$ leads us to find:  
\begin{equation}
\D{\sum_r}\var_r^{\mu} (\vec k)\var_r^{\nu} (\vec k') [\hat{a}_r(\vec k),\hat{a}_r^+(\vec k')]=-(\D{\sum_{\vec K}}\frac {K^{\mu}K^{\nu}}{{K_0}^2}\mid \vec K><\vec K\mid){\delta}_{\vec k \vec k'}
\end{equation} \label{47}
Of course, one may easily find that
$\left[ \hat{a}_r(\vec k),\hat{a}_r (\vec k') \right]=
\left[ \hat{a}_r^+(\vec k),\hat{a}_r^+ (\vec k') \right]=0$ and all other 
commutators vanish for $r\neq r'$. \\
A specific choice of polarization vectors in one given frame of reference 
often facilitates the interpretation. In the frame in which the photon
(anticipating the particle interpretation!) is moving along the third axis, we
have $K^{\mu}=(K,0,0,K)$, and
\begin{equation}
\var_0=(1,0,0,0), \var_1=(0,1,0,0), \var_2=(0,0,1,0), \var_3=(0,0,0,1)
\end{equation} \label{48}
Since our aim is to find a scalar, (the vacuum energy), which is a Lorentz invariant 
and also gauge independent quantity, this special choice of polarization 
vectors does not restrict the validity of discussion. 
By means of the above polarization vectors,
the following commutation relations can be found from (47)
\begin{equation}
 [\hat{a}_0(\vec k),\hat{a}_0^+(\vec k')]=-\hat 1{\delta}_{\vec k \vec k'}
\end{equation} \label{49}
\begin{equation}
 [\hat{a}_3(\vec k),\hat{a}_3^+(\vec k')]=+\hat 1{\delta}_{\vec k \vec k'}
\end{equation} \label{50}
\begin{equation}
 [\hat{a}_1(\vec k),\hat{a}_1^+(\vec k')]= [\hat{a}_2(\vec k),\hat{a}_2^+(\vec k')]=0
\end{equation} \label{51}
For "scalar" photons, we have the CR (49) just the 
same as in the usual quantum field theory of radiation. A 
remarkable point is the minus sign in the right hand side of (49)  
for which a comparison with the harmonic oscillator shows that, contrary to the 
usual interpretation, $\hat{a}_0(\vec k)$ and $\hat{a}_0^+ (\vec k)$ have the roles 
of creation and annihilation operators respectively.
The CR (50) is the same result
of the usual quantum field theory of radiation for longitudinal photons. The
difference appears in CRs (51).
In fact, (51) shows that the spectrum of the
eigenvalues of the operators $\hat{a}_i^+\hat{a}_i$ (i=1,2) is a continuous spectrum
of all real
values from zero to infinity because these operators are Hermitian
and positive definite. The continuity of the spectrum of these operators
is reasonable because we are working with free fields without any bound or
restriction unless for the special choice of reference frame whose effect is
clear in the CR (50).
Now, we are ready to find the expectation value of the Hamiltonian
for the
ground state i.e. vacuum.
The Hamiltonian is 
\begin{equation}
\hat{H}=\int d^3 x (\hat{\pi}_{\mu} \dot{A}^{\mu} -\hat{L})
\end{equation} \label{52}
Using (35)-(38) and (52) leads to 
\begin{equation}
\hat{H}=\D{\sum_r} \D{\sum_{\vec k}} \left[ \frac{\h \omega}{2} \right] \xi_r 
\left [\hat{a}_r^+
 (\vec k) \hat{a}_r (\vec k)+\hat{a}_r (\vec k)\hat{a}_r^+(\vec k) \right]
\end{equation} \label{53}
Substitution of (49), (50), and (51) into the relation  (53) leads to the following 
expression for the Hamiltonian
\begin{equation}
\hat{H}=\sum_{\vec k}(\h \omega) \left[ 
\hat{a}_1^+ (\vec k) \hat{a}_1 
(\vec k)+\hat{a}_2^+(\vec k)\hat{a}_2(\vec k)+\hat{a}_3^+(\vec k) \hat{a}_3(\vec k)- 
\hat{a}_0(\vec k)\hat{a}_0^+(\vec k)\right]
\end{equation} \label{54}
Since for the operators $\hat{a}_1^+\hat{a}_1$ and
$\hat{a}_2^+\hat{a}_2$ the minumum possible value of their
(continuous) spectrum is zero and since for scalar photons $\hat{a}_0^+(\vec k)$ has the role
of annihilation operator then we have
\begin{equation}
\langle 0|\hat{H}|0\rangle = 0
\end{equation} \label{55}
\noindent This achivement has been obtained without imposing the  normal ordering condition
which is usually used for removing the infinity of the vacuum energy.
For any case whose absolute value of energy is not relevant, normal ordering
will appear to be
adequate. But when the absolute value of energy is concerned, particularly 
in calculating the cosmological constant, normal ordering does not seem to be
reasonable.
A serious doubt about the above expression for the Hamiltonian is that for 
scalar particles one may deal with a state of negative infinite energy and 
we should have a satisfactory explanation for this problem (see Appendix -B). 
\subsection*{4. Conclusion}
Without imposing any condition, it turnes out 
that the free EM fields vacuum energy is zero and this can be 
considered as the solution of vacuum catastrophe [4]. It may be also 
considered as a step toward managing the cosmological constant problem 
[5].
Although there are many works on the solution of the cosmological constant problem 
e.g. supersymmetry [6], quantum cosmology [7-10], supergravity [11-13] 
and so on [14]; in all of them
there are either physically unknown assumptions and principles or
mathematical difficulties such as unmeasurability of the path integral used in 
quantum cosmology. But in this treatment the cosmological 
constant problem turns out to have probably a simple and physically reasonable solution at least for the 
contribution of EM fields to the vacuum state.\\
An important question which may raise to mind is that whether our result is in 
challenge with the Casimir effect [15]? The answer is no because
in this study we have dealt with the free electromagnetic fields Lagrangian
density,but in the case of the Casimir effect one should enter the effect of 
boundary conditions. This, however, requires a detailed and independent study.\\
\noindent {\bf Appendix -A} \\
In order to verify the microcausality condition, we will find the covariant 
commutator of the fields at two arbitrary points at first and then show that 
the result vanishes when the points have space-like separation. 
For the points $x:(ct, \vec x)$ and $x':(ct',\vec x')$, the covariant commutation 
relation
between $\hat{A}^{\mu} (x)$ and $\hat{A}^{\nu} (x')$ is 
\begin{equation}
\begin{array}{l} 
[\hat{A}^{\mu} (x), \hat{A}^{\nu} (x')]= 
\D{\sum_{r}} \D{\sum_{r'}} \D{\sum_{\vec k}}  \D{\sum_{\vec k'}} 
\left(\frac{ \h c^2}{2V \omega} \right)^{\frac{1}{2}}
\left(\frac{ \h c^2}{2V \omega'} \right)^{\frac{1}{2}}
\var^\mu_r(\vec k) \var_{r'}^{\nu}(\vec k') \\
\left [ \hat{a}_r(\vec k) e^{-ikx}+ \hat{a}_r^+ (\vec k) e^{ikx}, \hat{a}_{r'} 
(\vec k') e^{-ik' x'} +\hat{a}_{r'}^+(\vec k') e^{ik' x'} \right]
\end{array}
\end{equation} \label{56}
Taking the limit $V\rightarrow \infty$ we must 
substitue
$\D{\sum_{\vec k}} \frac{1}{V}$ by $\frac{1}{(2\pi)^3} \int d^3 k$ and (56) becomes
\begin{equation}
[\hat{A}^{\mu} (x), \hat{A}^{\nu} (x') ]=\frac{\h c^2}{2(2\pi)^3} 
\sum_r \var^{\mu}_r \var^{\nu}_r [\hat{a}_r, \hat{a}_r^+] \int 
\frac{ e^{-ik(x-x')}-e^{ik(x-x')}}{\omega} d^3 k
\end{equation} \label{57}
But, the integral $\int \frac{e^{-ik (x-x')}-e^{ik (x-x')}}{\omega} d^3 k$ is 
the  $\Delta$-function up to a multipilicative constant.  
$\Delta(x-x')$ is a Lorentz invariant function and $\Delta(\vec x-\vec x',0)=0$. 
Therefore $\Delta(x-x')$ vanishes for $(x-x')^2<0$ and the final result 
\begin{equation}
[\hat{A}^{\mu} (x), \hat{A}^{\nu} (x')]=0, for (x-x')^2<0
\end{equation} \label{58}
\noindent  is the microcausality condition.\\

\noindent {\bf Appendix -B}\\
In this work, as in the usual quantization of the free EM fields, we cannot 
simply take the Lorentz condition as an operator identity because it is
incompatible with the covariant form of the commutation relation (43). This problem may be resolved by replacing the Lorentz condition with the following weaker condition
\begin{equation}
<\Psi \mid \partial_{\mu}A^{\mu} \mid \Psi> = 0
\end{equation} \label{59}
\noindent where $\mid \Psi>$ is the physical state-vector of the system.
This ensures that the Lorentz condition and hence Maxwell's equations hold
as the classical limit of the theory.
In order to understand the meaning of the above subsidiary condition, we express
it in momentum space. On substituting the explicit form of $A_{\mu}(x)$, we
obtain
\begin{equation}
<\psi \mid (\hat {a_0} -\hat {a_3})e^{-ikx} +(\hat {a_0}^+ -\hat {a_3}^+)e^{+ikx} \mid \Psi> = 0
\end{equation}\label{60}
\noindent for which we have used the choice mentioned in the text where ${\var ^{\mu}}_r$ 
is orthogonal to $k_{\mu}$ for $r=1,2$.
Now, let define the following operators
\begin{equation}
\hat {A_0} = \hat {a_0}e^{-ikx}\\
\hat {A_3} = \hat {a_3}e^{-ikx}
\end{equation}\label{61}
\noindent Since here $\hat {A_0}^+$ is an annihiliation operator, at point x, 
if we demand to have the 
following condition
\begin{equation}
(\hat {A_0}^+ - \hat {A_3}) \mid \Psi> = 0
\end{equation} \label{62}
then, the above expectation value for Lorentz condition will be automatically 
satisfied.
Thus, the action of scalar particles compensates exactly that of the longitudinal
ones and we can always work in a gauge for which we deal with an appropriate admixture
of scalar and longitudinal photons not to have an infinite negative energy.

\end{document}